## Original Article

## Are Paralysed Chondrocytes Really Dying?

Ahmed Y. A.[1], Tatarczuch L.[2], El-Hafez A.[3], Zayed A. E.[3], Davies H. M.[2] and Mackie E. J.[2]

*[1]Faculty of Veterinary Science, South Valley University, [2]School of Veterinary Medicine, University of Melbourne, Australia, [3]Faculty of Veterinary Medicine, Assiut University.*

### ABSTRACT

**Introduction:** Paralysed chondrocytes are cells with uncommon morphology that recently have been seen among the non-hypertrophic chondrocytes of rabbit growth cartilage. Paralysed chondrocytes characterised by condensed nucleus and well developed rough endoplasmic reticulum and Golgi apparatous and numerous secretory vesicles. Because, in some examples, the cytoplasm of these cells appeared empty except from few organelles and the nucleus was condensed, paralysed chondrocytes were considered as an example of physiologically dying cells. Studying of paralysed chondrocytes is difficult due to the absence of an in vitro model in which the paralysed chondrocytes could be induced.

**Aim of the Work:** The aims of the current study were to establish a system of culture for induction of paralysed chondrocytes and to investigate if these cells are really dying.

**Results:** Chondrocytes were isolated from the growth cartilage of fetal equines, centrifuged and cultured as pellets in either 10% fetal calf serum or 10% horse serum for 28 days and processed for light and electron microscopy. Different cell types were counted and expressed as a percentage to the total cell number. Growth kinetics including the pellet weight and thickness and the cellular density were evaluated.

After 7 days in culture, paralysed chondrocytes with similar morphology to those described in the rabbit growth cartilage could be identified in pellets in each serum type, however, the proportion of the cells was different. In pellet cultured with 10% fetal calf serum, more than 50% of the cells were paralysed chondrocytes but in 10% horse serum, less than 10% of cells were of paralysed type. At day 14, about 50% of the cells in pellets cultured in either serum type differentiated into hypertrophic dark chondrocytes and the proportion of paralysed cells was markedly decreased. After 21 days in each culture, more than 70% of the cells were hypertrophic dark chondrocytes and no paralysed chondrocytes could be observed. Interestingly, the pellets in both serum types showed similar growth kinetics.

**Conclusion:** The paralysed chondrocytes may be not dying and they likely to be an immature form of hypertrophic dark chondrocytes. It is better to use the term immature dark chondrocytes instead of paralysed cells. This culture system will be useful for further molecular studies on paralysed chondrocytes and to explore the functions of these cells.

## INTRODUCTION

Longitudinal growth of the skeleton results from endochondral ossification. During this process, a cartilage model is formed by the condensation of the embryonic undifferentiated mesenchymal cells, which is replaced by bone and bone marrow. Three centers of ossification can be seen in a developing long bone: A primary ossification centre (POC) appears in the middle of the cartilage model, from which ossification extends towards both ends and then a secondary ossification centre (SOC) appears in each epiphysis[1]. The cartilage that remains between the POC and SOC is known as the physeal growth cartilage (PGC) or growth plate. The PGC participates in expansion of the POC and results in longitudinal bone growth. Underneath the permanent articular cartilage lies the articular epiphyseal growth cartilage (AEGC), which participates in expansion of SOC and results in enlargement and modeling of the epiphysis[2]. Growth cartilage is formed of chondrocytes embedded within the extracellular matrix (ECM), which is composed mainly of proteoglycans and collagens[3]. Appearance of growth cartilage varies with the stage of development and between PGC and AEGC, but chondrocytes are always present in three zones; resting, proliferative and hypertrophic zones[4,5,1]. Resting chondrocytes are small rounded and randomly distributed throughout the surrounding ECM and they are considered as stem cells responsible for generation of new proliferative chondrocytes[6,7]. Proliferative chondrocytes are flattened cells undergo mitotic division and arranged in columns parallel to the long axis of the bone. Hypertrophic chondrocytes are large rounded or oval cells and characterised by expression of collagen type X[8] and a high level of alkaline phosphatase[9].

Early literatures have reported that there are two morphologically different types of hypertrophic chondrocytes, dark and light[10-12], which have been described as different forms of one cell population[12]. More recently, we have confirmed that dark and light hypertrophic chondrocytes are two morphologically different populations of cells and they are also different in the patterns of their gene expression[13]. Furthermore, we have shown that dark and light hypertrophic chondrocytes die by different non-apoptotic modes of physiological cell death[13].

Roach and Clarke[14] introduced the term 'cell paralysis' to describe an additional type of chondrocytes with unusual morphology that were present among non-hypertrophic chondrocytes in the epiphysis of rabbit growth cartilage and they re-described them as another cell type in another study[15]. The paralysed cells were characterised by irregular condensed nucleus, strands of rough endoplasmic reticulum (RER) and disappearance of cytoplasmic organelles except for Golgi apparatus and some secretary vesicles. Depending upon this morphological description, the authors defined paralysed chondrocytes as a form of physiologically dying immature chondrocytes.

Understanding different forms of PCD (Physiological Cell Death) was difficult due to the lack of an in vitro system in which the isolated chondrocytes can be induced to undergo the same modes of death as seen in vivo. To study different forms of PCD of hypertrophic chondrocytes we have recently developed a system of culture, based on the pellet culture system introduced initially by Kato et al.[16], in which hypertrophic dying dark and light chondrocytes could be replicated[3]. In this culture system we found that the fetal but not the postnatal equine chondrocytes[17] behave in a similar way to those in growth cartilage in vivo. The aims of the current study were to examine the possibility of using such chondrocyte pellet culture system to induce paralysed chondrocytes in vitro and to understand if these cells are really dying.

## MATERIALS AND METHODS

### Reagents:

DMEM, gentamicin, amphotericin B, l-glutamine, fetal calf serum (FCS) and horse serum (HS) were obtained from Invitrogen (Carlsbad, CA, USA). Collagenase A was from Roche Diagnostics (Basel, Switzerland). L-ascorbic acid was from Sigma–Aldrich (St. Louis, MO, USA).

### Chondrocyte isolation and culture:

Metacarpophalangeal and metatarsophalangeal joints were obtained from 3-month equine fetuses collected from the local abattoir in Melbourne, Australia; the gestational stage was estimated on the basis of crown:rump length[18]. These places have been chosen because they have minimal muscles and skin, decreasing the chance of contamination with other non-chondrocyte cell types. The skin, muscle and tendons were cut away and the joints were opened under aseptic conditions. The superficial half of the cartilage close to the surface (formed mostly of resting cells) was removed from the chondroepiphysis (the cartilaginous end of the long bone before formation of the SOC) and dissected into small pieces. Chondrocytes were isolated by digestion in 0.5% collagenase for 12 hours in an incubator at 37°C with 5% $CO_2$. Aliquots of 5 X $10^7$ cells in 1 ml DMEM containing gentamicin (50 μg/ml), amphotericin B (2.5 μg/ml), L-glutamine (300 μg/ml), L-ascorbic acid (50 μg/ml) and 10% fetal calf serum (FCS) were transferred into 15-ml polypropylene tubes. Cells were centrifuged for 5 minutes at 1000 rpm and incubated at 37°C with the caps of the tubes closed. Chondrocytes were cultured in the presence of either 10% FCS or 10% horse serum (HS). After 48 hours, the cultured cells formed white disc-shaped pellets that could be manipulated with forceps. Pellets were collected at 0, 7, 14, 21 and 28 days and processed for light and electron microscopy as described below.

### Phase contrast and electron microscopy of pellet culture:

Pellets were collected after specific time point of cultures and fixed in 2.5% glutaraldehyde/4% paraformaldehyde in 0.1 M cacodylate buffer (pH 7.4) for 24 hours at 4ºC, then postfixed in 1% osmium tetroxide/1.5% potassium ferrocyanide and embedded in Spur's resin. Semi-thin sections were stained with methylene blue for Phase contrast. Ultra-thin sections were stained with Uranyl acetate and Reynold's stained and examined with a transmission electron microscope (Philips 300).

### Cell counts:

Counts of different cell types were undertaken manually from semi-thin sections of pellets stained with methylene blue using a 100x under oil immersion lens of a light microscope. Paralysed chondrocytes, depending on the morphological definition of Roach and Clarke[14] that they are cells with cytoplasm nearly empty except from few organelles and following extensive comparison of light and electron micrographs of the same specimens, were counted and expressed as a percentage to the total cell number. Dark, light and resting chondrocytes, depending on the previously described morphological appearances[3], were counted and expressed as a percentage of total cell number. Results were presented as mean ± standard error (SE; n = 3). Statistical differences between groups were evaluated using a one way, ANOVA; P values < 0.05 were considered significant.

## RESULTS

### Induction of paralysed chondrocytes in vitro:

Chondrocytes isolated from the growth cartilage of fetal horses were cultured as pellets in 10% FCS or 10% HS for 28 days. At day 0, the freshly isolated chondrocytes were rounded and closely packed together with no evidence ECM (Fig. 1-B). Most of the cells showed the morphology of chondrocytes from the resting zone (Fig. 1-B) and (Fig. 2-D) and no paralysed chondrocytes could be seen. With time in culture, the isolated chondrocytes formed a typical piece of cartilage-like tissue grossly (Fig. 1-A) and histologically, they were embedded in their lacunae into abundant ECM (Fig. 1-D) and surrounded with perichondrium-like layer of flattened cells (Fig. 2-B).

Paralysed cells, similar to those described in previous studies[15] (condensed nucleus and nearly cytoplasm with many empty spaces), could be identified in pellets at day 7 of culture in  phase contrast (Fig. 1-C) and electron (Fig. 2-C) microscopic preparations, but disappeared completely at day 21 leaving no cell debris in the lacunae previously occupied with them. In addition, hypertrophic dark (Fig. 2-E) and light (Fig. 2-F) chondrocytes, with similar morphology to those previously described[3], were observed.

### Counting of different cell types in culture:

Counting different cells types (paralysed, resting, dark and light) was conduced on semi-thin sections of pellets from 3-month equine fetuses cultured in 10% FCS or HS for 28 days. It was found that at day 7, paralysed cells was more than 50% of the total cell number of pellets grown in 10% FCS, however in 10% HS, they were less than 10%. Paralysed chondrocytes significantly decreased at day 14 and a substantial proportion (about 50%) of cells was differentiated dark chondrocytes with few light chondrocytes. No paralysed chondrocytes were present in pellets cultured for more than 14 days. After 21 days in each serum type, more than 70% of the culture was hypertrophic dark chondrocytes (Fig. 3) no paralysed chondrocytes could be identified in this stage of differentiation.

### Estimation of the growth kinetics of pellet cultures in both serum:

The  effect of 10% HS on pellet growth was compared with that of 10% FCS. The growth kinetics of these pellets were evaluated by growth parameters such as pellet diameter, weight, thickness and cellular density. The pellet diameter remained unchanged at $5 \pm 1$ mm after 7 days regardless of the length of time in culture and whether they were cultured in HS or FCS. The pellet weight was $5.5 \pm 0.3$ mg (10 % FCS) and $4.5 \pm 0.5$ mg (10% HS) at day 7. Over 28 days in culture, the pellet weight reached $9.1 \pm 0.5$ mg (10% FCS) and $9.6 \pm 0.6$ mg (10% HS). No significant differences in pellet weight were found between the two types of serum at any time point of culture (Fig. 4-A). The pellet thickness significantly increased over the period of culture from $143 \pm 5$ μm (10% FCS) and $128 \pm 3.5$ μm (10% HS) at day 7 to $453 \pm 34$ μm (10% FCS) and $445 \pm 83$ μm (10% HS) at day 28.  No significant differences in pellet thickness were found between the two types of serum at any time point of culture (Fig. 4-B). The cellular density was $3.8 \pm 0.5$ cells/mm² (10% FCS) and $2.4 \pm 0.3$ cells/mm² (10% HS) at day 7 and decreased to $1.2 \pm 0.3$ cells/mm² (10% FCS) and $1.1 \pm 0.15$ cells/mm² (10% HS) at day 28. The cellular density was significantly lower in 10% HS than 10% FCS at day 7, however, no significant differences in the cellular density were present between the two types of serum after 7 days (Fig. 4-C).

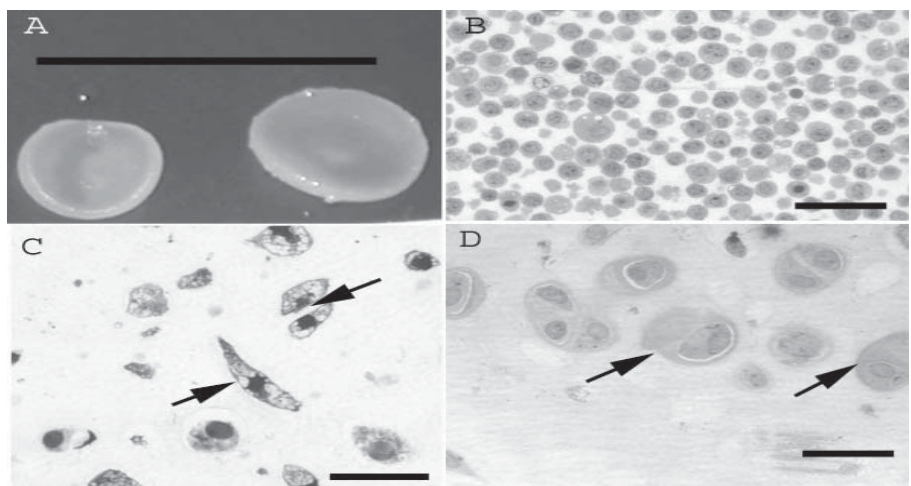

**Fig. 1:** Induction of paralysed chondrocytes in pellet culture system:
**A:** Gross appearance of the chondrocytes pellet culture. **B-D:** Phase contrast from the chondrocytes pellet culture at day 0 (B), 7 (C) and 14 (D). Arrows indicate paralysed chondrocytes in C and differentiated hypertrophic chondrocytes in D. Bars = 1 cm in A,  50  μm in B, 25 μm in C and 20 μm in D.

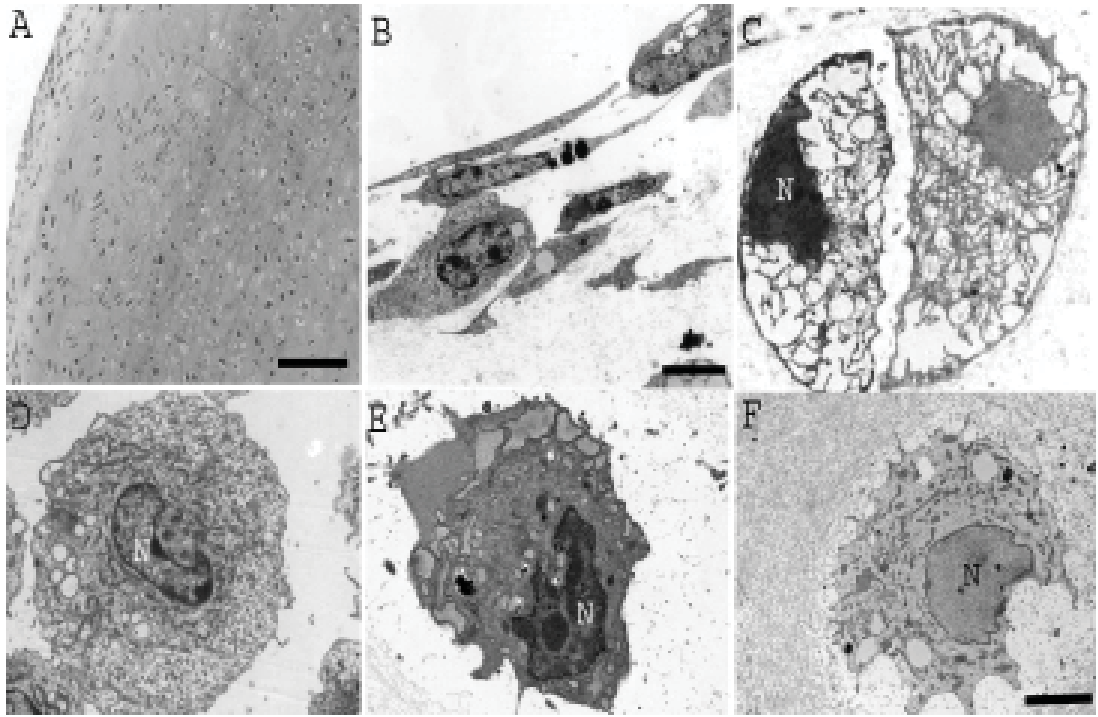

**Fig. 2:** Morphology of different cell types in pellet cultures:

Phase contrast **(A)** and electron micrographs **(B-F)** from pellet cultures showing cartilage like tissue appearance of the culture **(A)**, Perichondrium-like layer of flattened cells surrounding the cultures **(B)**, paralysed **(C)**, resting **(D)**, hypertrophic dark (E) and light (F) chondrocytes. Note the pattern of nuclear condensation (N). Parts C-F have the same magnification. Bars in A = 125 µm, in B = 2 µm and in F = 0.

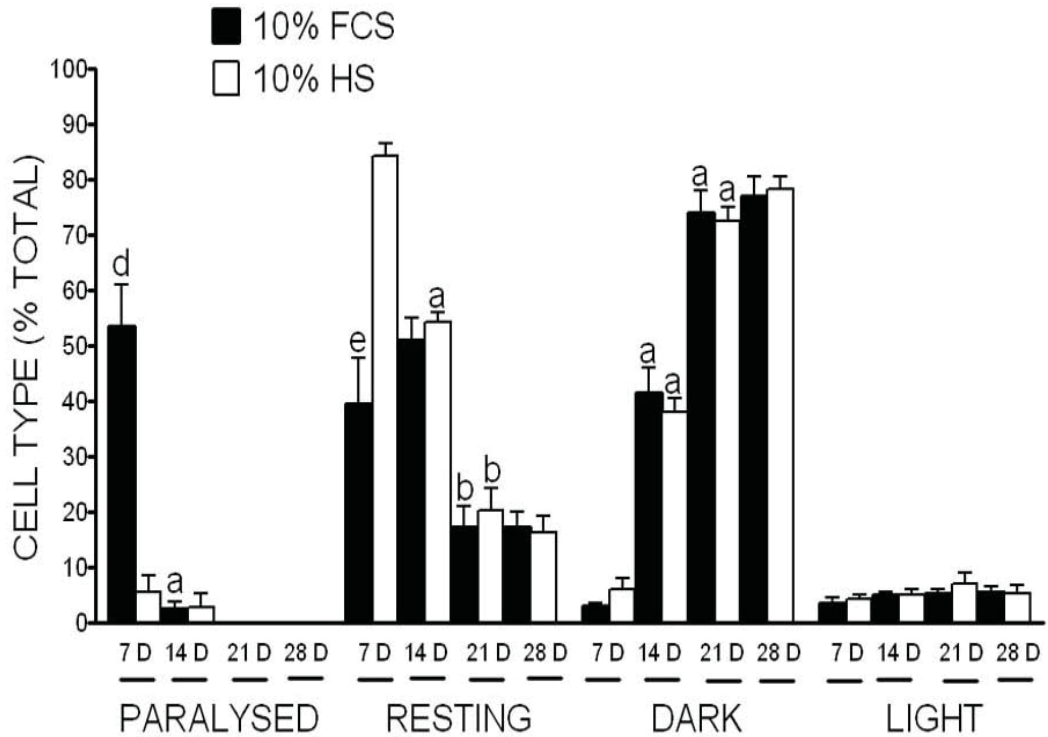

**Fig. 3:** Counting of different cell types in pellet cultures in FCS and HS:

Chondrocytes were isolated from growth cartilage of 3-month equine fetuses and cultured in the presence of either 10% FCS or 10% HS for up to 28 days. Semi-thin sections were taken from pellets cultured for 7 days (7 D), 14 days (14 D), 21 days (21 D) and 28 days (28 D). Paralysed, resting, dark or light chondrocytes were counted and expressed as a percentage of total cell number. Results are presented as mean ± SEM (n ≥ 3). Significant differences between each two successive time points in either serum type are indicated as follows: a (P < 0.001), b (P < 0.01), c (P < 0.05). A significant difference between serum types at any time point is indicated as d (P < 0.001), e (P < 0.01).

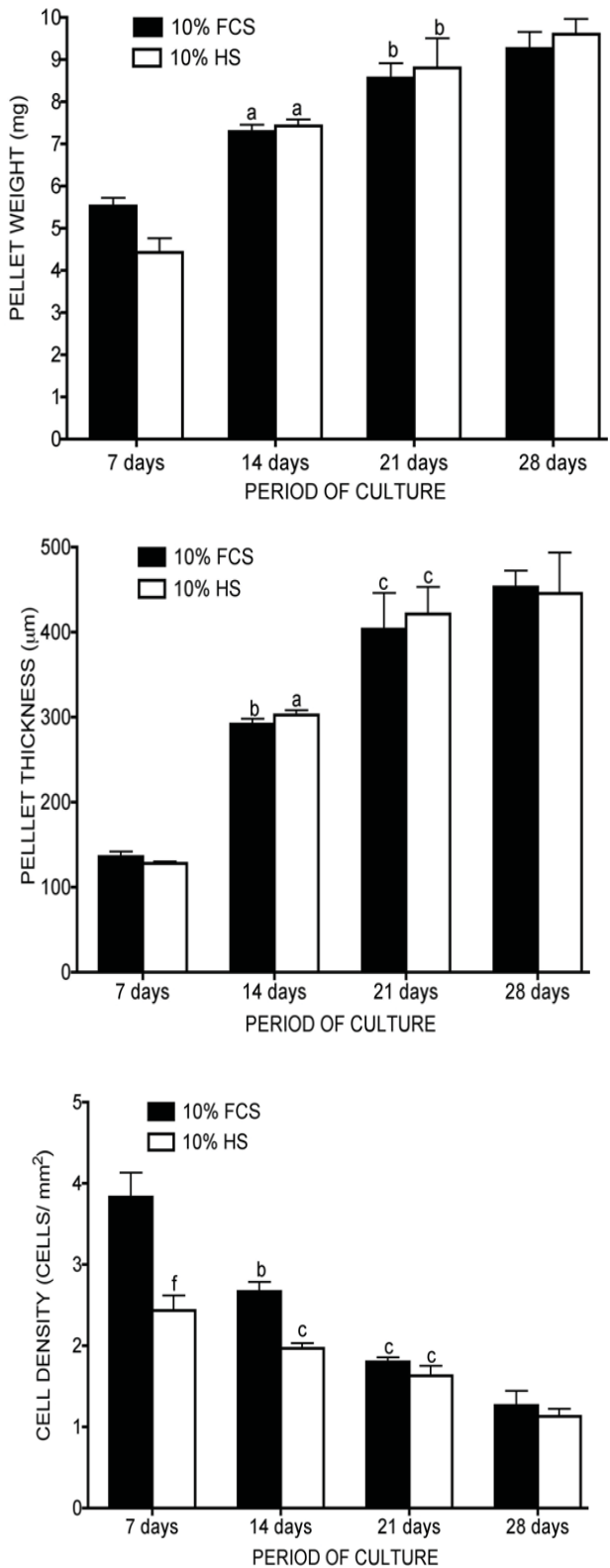

**Fig. 4:** Growth changes in pellet cultures in FCS and HS:

Chondrocytes were isolated from growth cartilage of 3-month equine fetuses and cultured as pellets in either 10 % FCS or 10 % HS and the pellets were collected after 7, 14, 21 and 28 days. The pellet weight (A), thickness (B) and cellular density (C) were measured. Data represent mean ± SEM (n ≥ 3). Significant differences are expressed as follows: a (P < 0.001), b (P < 0.01), c (P < 0.5) for comparison between each two successive time points in each serum type, f (P < 0.05) for comparison between serum types at day 7.

## DISCUSSION

The current study was undertaken with the aims first to induce paralysed chondrocytes in vitro and second to investigate if they are really dying cells. To achieve these aims, chondrocytes were isolated from 3-month equine fetuses and cultured as pellets in the presence of 10% FCS or HS for a period up to 28 days and the pellets were collected at time intervals and processed for phase contrast and electron microscopy. At day 0, no paralysed chondrocytes were seen, however, after 7 days in culture, paralysed chondrocytes could be identified in pellet cultures in both serum type suggesting that some chondrocytes are programmed to undergo this type of differentiation in culture.

Cell count of different cell types was undertaken. It was demonstrated that at day 7, there were significantly more paralysed chondrocytes in pellets grown in 10% FCS than those grown in 10% HS. These cells constituted 50% of the total cell number in 7-day pellets grown in 10% FCS. If the paralysed chondrocytes were dying cells as suggested[14], it would be likely that at 14 days a considerable amount of cell debris would be observed in the pellets and this is not the case.

Moreover, if the paralysed cells were dying, the pellets in FCS would not be expected to grow as rapidly as pellets in HS, which have comparatively a very few percentage of paralysed cells at day 7. In a part of our studies it was found such pellets grown in FCS with more than 50% paralysed chondrocytes at day 7 showed an increase in their growth rate during the 28 days of cultures to the same extent as pellets in HS did. Moreover, after 21 days in cultures, cells in pellets cultured in each serum types contained more than 70% of cells as differentiated hypertrophic dark chondrocytes suggesting that paralysed chondrocytes undergo dark cell differentiation.

From these in vitro findings, it is concluded that paralysed chondrocytes are not dying and they likely to be an immature form of hypertrophic dark chondrocytes. Thus we recommend using the term immature dark chondrocytes instead of paralysed cell. The culture system presented in this study may be useful for further studies on the difference in gene expression between paralysed chondrocytes and other types of chondrocytes and to better understand the role of paralysed chondrocytes in development of the growth cartilage.

### Acknowledgments:

This study was partially funded by Linkage Grant no. LPO348867 from the Australian Research Council and Racing Victoria Ltd. The authors thank Mr. Brendan Kehoe for assistance with obtaining the specimens and Ms. Su Toulson for assistance with the histological studies.

**هل الخلايا الغضروفية المشلولة هي خلايا ميتة؟**

ياسر عبد الجليل احمد [1]،تاتازروك[2]، إيناس احمد عبد الحافظ[3]، احمد الزهري زايد[3]،دافيز[3] و ماكي[3]

كلية العلوم البيطرية ، جامعة جنوب الوادي ، قنا ، مصر ، كلية الطب البيطري ، جامعة ملبورن ، استراليا ، وكلية الطب البيطري ، جامعة أسيوط ، أسيوط ، مصر .

| ملخص البحث |
| --- |

الخلايا الغضروفية المشلولة هي خلايا غير معهودة مورفولوجيا" ولكنه تم التعرف عليها ووصفها حديثا"  بين الخلايا الغضروفية الغير متضخمة في غضروف النمو في الأرانب. وتتميز هذه الخلايا بنواة مكثفة و غامقة و كذلك شبكة اندوبلازمية محببة و أجسام جولجي متطورين وأيضا" تتميز باحتوائها على العديد من الحويصلات الإفرازية . ولأنه في بعض الأمثلة،  بدا سيتوبلازم هذه الخلايا خاليا" إلا من عدد قليل من عضيبات الخلية. اعتبرت هذه الخلايا مثالا"" للموت الفيسيولوجي لخلايا الغضروف الغير متضخمة.

وكان من الصعب دراسة هذا النوع من الخلايا نظرا" لعدم وجود نظام لزراعة الأنسجة يمكن فيه تحفيز الخلايا الغضروفية المعزولة من غضروف النمو لتسلك مسلك الخلايا المشلولة في المعمل. لذا كانت أهداف الدراسة الحالية هي ابتكار نظام لزراعة الأنسجة يمكن فيه محاكاة هذه الخلايا بحيث يمكن دراستها في المعمل، و بالتالي التحقيق في ما إذا كانت الخلايا المشلولة هي خلايا ميتة.

لتحقيق هذين الهدفين تم عزل الخلايا الغضروفية الغير مميزه من غضروف النمو في أجنة الخيول و زرعها في أنابيب من البوليبروابلين في وسط يحتوي علي إما 10 % مصل عجل جنيني أو 10 % مصل حصان لمدة 28 يوم. تم تجميع الخلايا المزروعة بعد فترات منتظمة و تجهيزها للفحص بالمجهر الضوئي والمجهر الإلكتروني. و قد تم عد أنواع الخلايا المختلفة التي تتميز إليها الخلايا المعزولة بعد زراعتها و التعبير عنها كنسبة مئوية للعدد الكلي للخلايا . وكذلك تم حساب بعض التغيرات الديناميكية التي تطرأ عليها مثل الوزن والسمك و الكثافة الخلوية للغضروف الناشئ من زراعة هذه الخلايا.

بعد 7 أيام من زراعة هذه الخلايا، لوحظ تميزها إلي خلايا مشابهة تماما" للخلايا المشلولة التي تم وصفها في غضروف النمو في الأرانب في الدراسات السابقة. كما لوحظ أيضا" وجود نسبة كبيرة من هذه الخلايا (50%) في وجو مصل عجل جنيني بينما كانت هذه النسبة أقل من 10% في وجود مصل حصان. بعد اليوم الرابع عشر انخفضت نسبة الخلايا المشلولة في وجود أي من المصلين بشكل ملحوظ لتختفي تماما" بعد 21 يوما. وتميز أكثر من 70 % من الخلايا بعد 21 يوما" إلي الخلايا المتضخمة الغامقة. وكان مثيرا"" للاهتمام أن نجد أن وزن وسمك الغضروف الناشئ و كثافة الخلايا متساوي بعد 14، 21، 28 يوما" وجود أيا" من المصلين. من هذا أمكننا التعليق بأن الخلايا المشلولة ربما لا تكون خلايا ميتة ويحتمل أن تكون خلايا غير ناضجة تتميز مع الوقت إلي خلايا متضخمة غامقة. لذلك نوصي باستخدام المصطلح الخلايا الغضروفية الغامقة الغير ناضجة بدلا" من مصطلح الخلايا المشلولة.  نظام زراعة الأنسجة الذي تم استخدامه سيكون مفيدا" لإجراء المزيد من الدراسات حول البيولوجيا الجزيئية للخلايا المشلولة، واستكشاف وظائف هذه الخلايا.